\documentclass[prl]{revtex4}
\input epsf
\usepackage{graphics}
\usepackage{graphicx}
\usepackage{float}
\usepackage{amsmath}
\usepackage[francais,english]{babel}
 
\newcommand{\beq}{\begin{equation}} \newcommand{\eeq}{\end{equation}}
\long\def\omitt#1 {}

\begin{document}

\title{\bf Stability of giant vortices in quantum liquids}

\author{Christophe Josserand}
\affiliation{Laboratoire de Mod\'elisation en M\'ecanique,\\
CNRS UMR 7607, 8 rue du Capitaine Scott,\\ 75015 Paris-France}

\begin{abstract}
We show how giant vortices can be stabilized in strong external potential 
Bose-Einstein condensates. We illustrate the formation of these vortices 
thanks to the relaxation Ginzburg-Landau dynamics for two typical potentials in
two spatial dimensions.
The giant vortex stability is studied for the particular case of the rotating 
cylindrical hard wall. The minimization of the perturbed energy is simplified
into a one dimensional relaxation dynamics. The giant vortices can be 
stabilized only in a finite frequency range. Finally we obtain a curve for the
minimum frequency needed to observe a giant vortex for a given nonlinearity.\\
{\bf Keywords:} Bose-Einstein condensates, quantized vortex, multiple charge, 
energy minimization.
\end{abstract}

\maketitle

{\bf Quantized vortices are characteristics of quantum systems. They appear 
for instance when a rotation is imposed to a quantum liquid. We study here the
conditions under which vortices with multiple charges can be observed. It is
in general not possible since the energy of an array of single vortices is
below the energy of a multiple vortex of same total charge. However an 
opposite trend appears for the energy if the system is confined. This 
in particular the case for the recent achieved Bose-Einstein condensates where
an optical trap maintains the gas in a small volume.
We show that if the trap potential is strong enough the 
multiple charge vortex becomes stable. The linear stability of these 
giant vortices is performed using one dimensional minimization techniques.
}
\section{Introduction}

The recent achievement of Bose-Einstein condensates 
(BEC)\cite{BEC,And95,Dav95} has strongly renewed and restimulated the
 numerous works and theories on quantum gases/liquids. By their lower 
densities, their larger healing lengthes, the BEC present important advantages
compared to the traditional Helium superfluids, regardless the fundamental 
interest of the Bose-Einstein transition itself. One of the 
striking properties of these quantum
gases/liquids is the quantization of the vorticity\cite{He4}. 
It manifests through the formation of vortices whose circulation is 
quantified in units of $\hbar/m$, where
$m$ is the mass of an atom of the gas and $\hbar$ is the Planck constant. 
Such vortices have been observed in BEC by two different procedures these last 
years\cite{dal,kett1,kett2}. Firstly\cite{dal} by rotating the BEC through 
laser 
techniques. This experiment is the analogy of the famous rotating bucket of
superfluid helium where the first direct observation of quantized vortices
was obtained\cite{pack}. The vortices appear suddenly as the rotation 
frequency increases, either one by one, or by group, depending on the 
experimental procedure. The other experiment\cite{kett1,kett2} exploits the 
dynamical transition to vortex shedding of an accelerated superflow\cite{fpr}.
Experimentally, to achieve the BE condensation the atoms of the condensate 
have to be placed in an optical trap which creates an effective (mostly 
harmonic) attractive potential. Thus, in the above experiments only single 
charged vortices were observed. No multicharged vortices (called giant vortex 
of charge $q$ or $q$-vortex later on) were detected. However, by imposing 
specific initial states, it has been possible to observe metastable long-lived
giant vortices experimentally \cite{cor03}. It is in fact well known that 
multicharged vortices are always dynamically and thermodynamically unstable 
in homogenous and infinitely large quantum liquids\cite{aronson}. When the 
system is submitted to a trap potential, the situation becomes more tricky 
since the potential encourages vortex merging while the vortices try to 
expand the condensate. For BEC in harmonic trap, variational stability 
analysis suggests indeed the existence
of a critical frequency above which the giant vortex should be energetically
favoured to the single vortices lattice\cite{CaDu99}. But this
frequency is found higher than the frequency of the quadratic
trapping potential. At such speed, the condensate is unstable since
the centrifugal force overcomes the trap confinement. This explains
why no stable giant vortices have been observed until now in rotating BEC,
where only harmonic trap were experimentally considered. This peculiar 
property of harmonic confinement does not actually prevent the stability 
of giant vortices for stronger trap potentials.
In fact different groups have shown numerically and argued
analytically that giant vortices are stable for high enough rotation 
if the trap potential is taken stronger than 
harmonic\cite{Lu01,Ueda02,FiBa01}. Numerical simulations have exhibited 
these giant vortices and some coexistence diagramms were even
obtained\cite{Lu01,Ueda02}, although no clear 
stability arguments were presented. In the other work\cite{FiBa01} the 
stability analysis was only performed in the framework of Wigner-Seitz 
approximation of a vortex lattice. Adequate strong trap potentials can be 
experimentally achieved thanks for example to Laguerre-Gaussian laser 
beams\cite{Kuga97} but the observation of stable giant vortices remains still
to be done.

In this note, we address the stability analysis of the giant vortices for the
rotating cylinder.
We use numerical simulations for two-dimensional systems subject to 
rotation ${\bf \Omega}=\Omega {\bf z}$, assuming translational invariance in
the $z$-direction. The giant vortex formation is shown for illustration
in a rotating BEC with quartic potential but we restrict thereafter our
treatment to the case of an infinite cylindrical potential.

\section{The model}

The dynamics of a quantum Bose gas in the rotating frame in two dimensions
is described through the dimensionless Gross-Pitaevski\v{\i} (G-P) 
equation\cite{Gross,Pit,String}:

\begin{equation}
i \partial_t \psi=\left( -\frac{1}{2} \nabla^2+V(r)+M |\psi|^2 +i \Omega
\partial_\theta \right) \psi 
\label{GP}
\end{equation}
\noindent
where $\psi({\bf r},t)$ is the condensate wave function normalized to unity: 
$\int d{\bf r} |\psi|^2=1$. ${\bf r}$ is the
position vector $(x,y)$, $r=|{\bf r}|$ and $\theta$ the corresponding 
cylindrical coordinates. The nonlinear strength $M=4 \pi N  a$ is related to 
the product of the 
particle number per unit length $N$ with the s-wave scattering length $a$.
This dimensionless equation has been in fact deduced from the usual model using
a typical length scale $d$ that is defined by the trap properties. Consequently
the length, energy and time in equation (\ref{GP}) are given in units of $d$, 
$\hbar^2/(md^2)$ and $md^2/\hbar$ ($\hbar$ the Planck constant and $m$ the
particle mass). The usual harmonic potential
$ \frac{1}{2}m\omega_t^2r^2$ leads to
$V(r)=\frac{1}{2}r^2$ with $d=\sqrt{\hbar/(m\omega_t)}$. Below we 
restrict our work to only two traps geometry. Firstly for illustration a 
quartic potential 
$\frac{m^2 \omega_t^3 r^4}{\hbar} $, so that $V(r)=\frac{1}{4}r^4$ and 
$d=\sqrt{\hbar/(m\omega_t)}$ again. This potential has a smooth behavior and 
describes a stronger trap potential than the usual harmonic potential of the 
experiments. The other potential models a quantum liquid entrapped in a
cylindrical bucket of radius $L$. With $d=L$ the potential is described by
$V(r)=0$ if $r<1$ and $V(r)=\infty$ otherwise. It corresponds to the well-known
experiment on superfluid Helium where quantized vortices have been first 
vizualized\cite{pack}. Note that equation (\ref{GP}) admits only two control
parameters for a given potential: the interaction coefficient
$M$ and the angular frequency $\Omega$. The dynamics derives
from a free energy $F$ through:
$ \imath \partial_t \psi=\frac{\delta F}{\delta \psi^*} $ with:

\begin{equation}
F= \int d{\bf r} \left( \frac{1}{2} |\nabla \psi |^2 +i \Omega \psi^* 
\partial_\theta \psi+V(r)+\frac{M}{2}|\psi|^4 \right)
\label{fener}
\end{equation}

Stable equilibrium solutions of Eq. (\ref{GP}) correspond to local minima of 
$F$. The ground state is determined among the stable equilibrium solution as 
the one with the lowest energy . Stable equilibrium states of the condensate 
can be reached numerically through the dissipative dynamics of the 
Ginzburg-Landau equation. Formally it corresponds to an imaginary time 
evolution of the GP equation (\ref{GP}):

\begin{equation}
\partial_t \psi=-\left( -\frac{1}{2} \nabla^2+4\pi N a |\psi|^2-\mu(N,\Omega)
+V(r)+i \Omega \partial_\theta \right) \psi 
\label{GL}
\end{equation}

We have introduced the chemical potential $\mu$ as the Lagrangian 
multiplier ensuring that the norm of $\psi$ is unity.
Numerical simulations have been performed in two space dimensions up to 
$256 \times 256$ grid points. We have checked that numerical precision is not
significantly affected by higher spatial discretizations for our range of 
parameters. Both Fast Fourier Transform (FFT) or 
finite difference methods have been used for spatial derivatives and again
no dramatic changes were noticed between these schemes.
The relative error in the normalization of the wave function was controlled to 
be below $10^{-3}$ and the precision for $\mu$ is better than $10^{-3}$.

\section{Giant vortex formation}

To illustrate the formation of giant vortices, we follow numerically
an equilibrium solution as $\Omega$ evolves. We start 
at $\Omega=0$ with the ground state (vortex free solution) and we increase
suddenly $\Omega$ of a small increment $\delta \Omega$. We retrieve an
equilibrium state through the imaginary time dynamics (\ref{GL}) and the 
procedure is then successively reproduced. Both the quartic and the hard-wall
potentials are considered.

\begin{figure}
\centerline{\includegraphics[width=16cm]{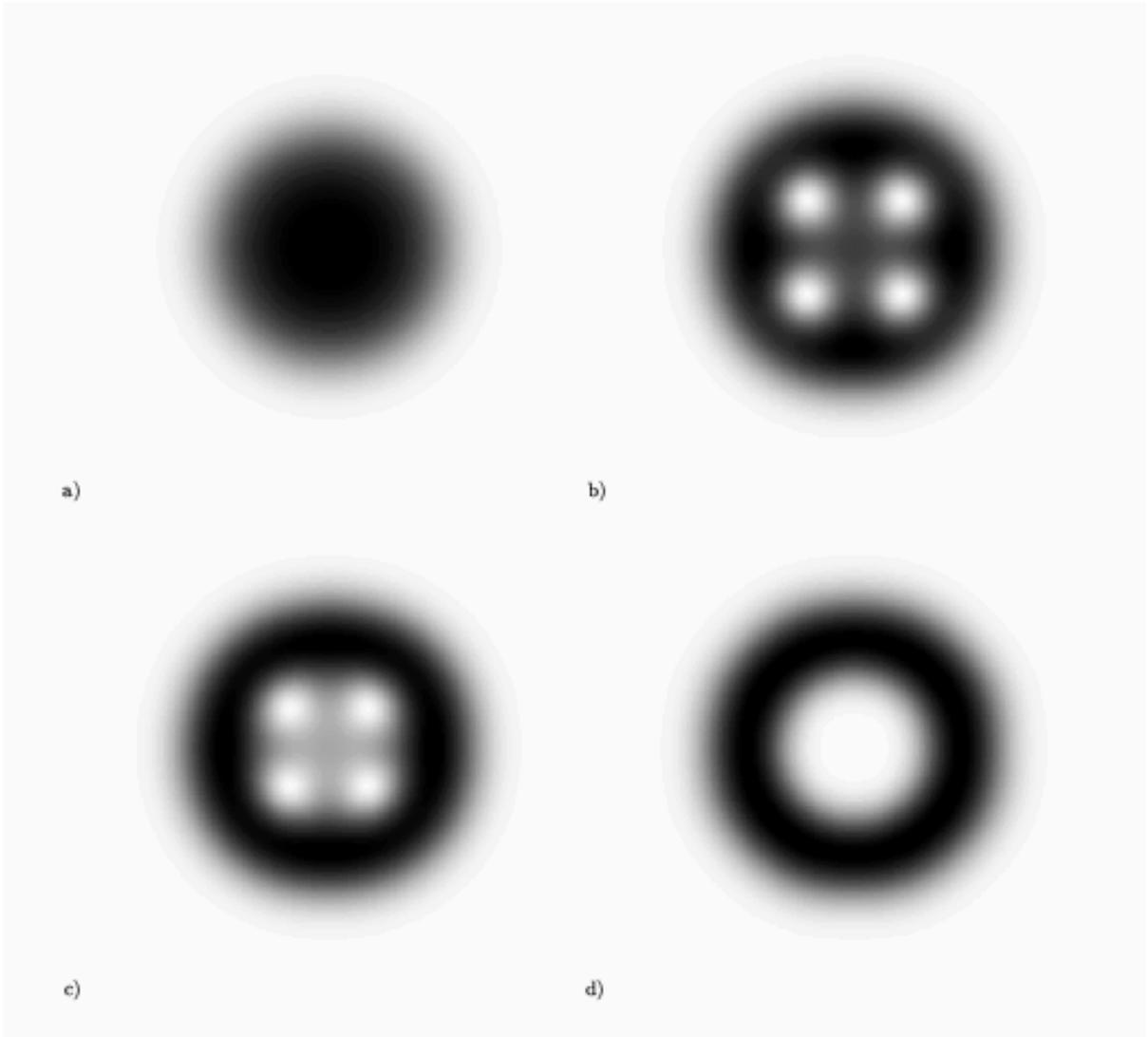}}
\caption{\protect\small
Quartic trap potential: density plots $|\psi|^2$ for 
$N \cdot a=3$ at $\Omega=$
a) $0.$, b) $5.4$, c) $6.6$, and d) $7.2$. The grid size
is $ 128 \times 128$ and a pseudo-spectral method is used. The darker 
the color, the higher the density so that the vortices appear as white hole
in the condensate.
\label{quar0}}
\end{figure}

We first present the results for the quartic potential with $M=12 \pi$ 
(figures \ref{quar0}). Figure (\ref{quar0}.a) shows the ground state
solution, vortex-free at $\Omega=0$. This solution survives without any
change until $\Omega=5.4$ where suddenly four vortices enter into the cloud
(figure \ref{quar0}.b). They are well separated and form a 
square lattice. We observe also a spatial expansion of the cloud due to
centrifugal effects.
As $\Omega$ increases further on(figure \ref{quar0} c)), 
the four vortices approach slowly the center of the trap. In this case 
the full merging of the four vortices is attained at  $\Omega=7.2$ (figure 
\ref{quar0} d)), just before new vortices enter the cloud (at $\Omega=7.35$).

\begin{figure}
\centerline{\includegraphics[width=16cm]{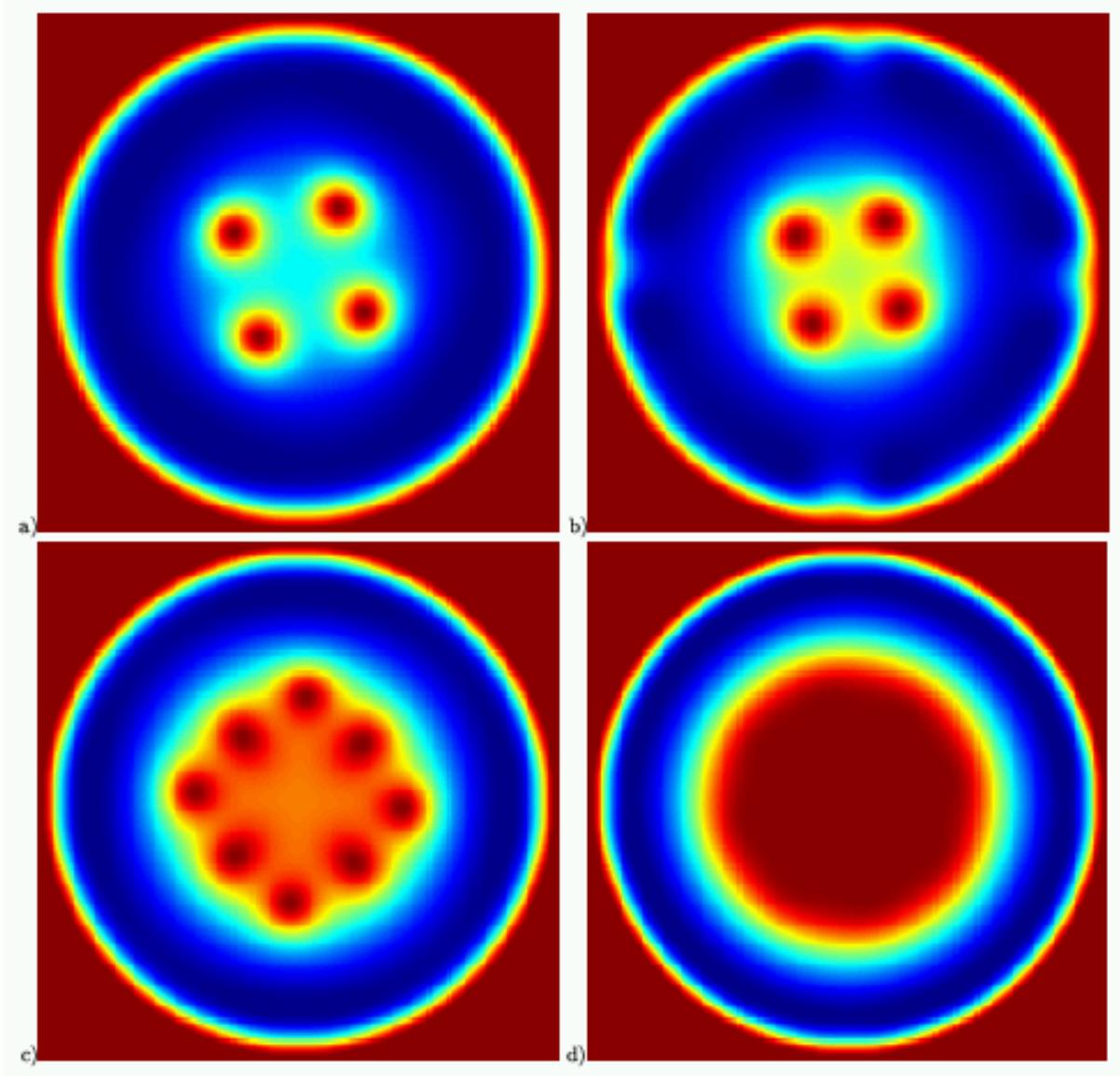}}
\caption{\protect\small Density plots for $M=300$, for $\Omega=$
a) $14.8$, b) $18.$, c) $20.8$, and d) $26.2$. The grid size
is $ 250 \times 250$ and the increment of the angular frequency is $0.2$. 
The color scale goes from blue for zero density to red for the maximum density. The vortices appear as blue holes on the
density field.
\label{snap0}}
\end{figure}

A similar situation is observed for a superfluid in a cylindrical trap
although in that case the condensate cannot expand when vortices are nucleated.
Indeed we account for the infinite hard wall potential by imposing as
boundary conditions $\psi=0$ for $r \ge 1$.
Figure (\ref{snap0}) shows the density $|\psi|^2$ of the solution of equation
(\ref{GL}) as $\Omega$ increases for $ M=300 $. The nucleation of four 
vortices occurs at $\Omega=14.8$ (figure (\ref{snap0} a)) from 
the vortex free solution. Further increments of
$\Omega$ lead to the convergence of the four vortices towards
the center of the trap(figure (\ref{snap0} b)). But a different scenario arises here compared
to the dynamics observed for the quartic potential. Indeed, before the four
vortices collapse in a single $4$-vortex, four new vortices 
are nucleated in the cloud at $\Omega=20.8$(figure (\ref{snap0} c)). 
The condensate is now composed of a lattice of eight well separated vortices. 
Then new increases of $\Omega$ lead 
eventually to the merging into a giant $8$-vortex at $\Omega=26.2$ 
(figure (\ref{snap0} d)). Above this rotation, when
new vortices are nucleated, they are immediatly absorbed by the giant vortex.
This persistence of giant vortices for increasing frequencies has been in
fact observed for all the numerical simulations that we have performed.

\section{Stability analysis}

By sake of simplicity, the stability analysis is performed for the 
cylindrical wall trap potential. We want to address the conditions under which 
conditions a $q$-vortex is a local minimum of the free energy (\ref{fener}).
We determine firstly the giant $q$-vortex state by seeking solutions of the 
form:

$$ \psi_q^0({\bf r})=f_q(r)e^{i q\theta} $$

The trivial symetry of the problem $q \rightarrow -q$ with
$\Omega \rightarrow -\Omega$ makes us only consider later on the
case $q>0$.
From the free-energy principle, to be an equilibrium 
solution $f_q$ has to satisfy:

\begin{equation}
\frac{1}{2} \left(f_q''+\frac{f_q'}{r}-\frac{q^2}{r^2}f_q 
\right) - M f_q^3 + (\mu_q(M,\Omega)+q \cdot \Omega ) f_q=0 
\label{eqaxi}
\end{equation}
\noindent
$\mu_q(M,\Omega)$ is introduced again as the Lagrangian multiplier 
for the norm constraint $ <\psi_q^0|\psi_q^0>=<f_q |f_q>=
2\pi \int rdr f_q^2=1 $, where $<\cdot |\cdot>$
stands for the usual scalar product. The external potential term is replaced
by the boundary condition $\psi=0$ for $r \ge 1$ so that this potential term
is everywhere zero in the equation. We observe from the preceeding 
equation (\ref{eqaxi}) that the giant vortex solution is independant 
of $\Omega$ which only
intervenes in the coefficient $ \mu_q(M,\Omega)=\mu_q^0(M)-q \cdot \Omega$.
These solutions are thus equilibrium states for all frequencies $\Omega$ 
regardless their stability. The free energy of the vortex writes:

\begin{equation}
F_q= 2\pi \int r \cdot dr \left( \frac{1}{2}((f_q')^2+\frac{q^2}{r^2}f_q^2) 
- q\Omega f_q^2 +\frac{M}{2}f_q^4 \right) =T_q(M)+U_q(M)-
q \cdot \Omega
\label{ener1d}
\end{equation}

\begin{figure}
\centerline{a)\epsfxsize=8truecm \epsfbox{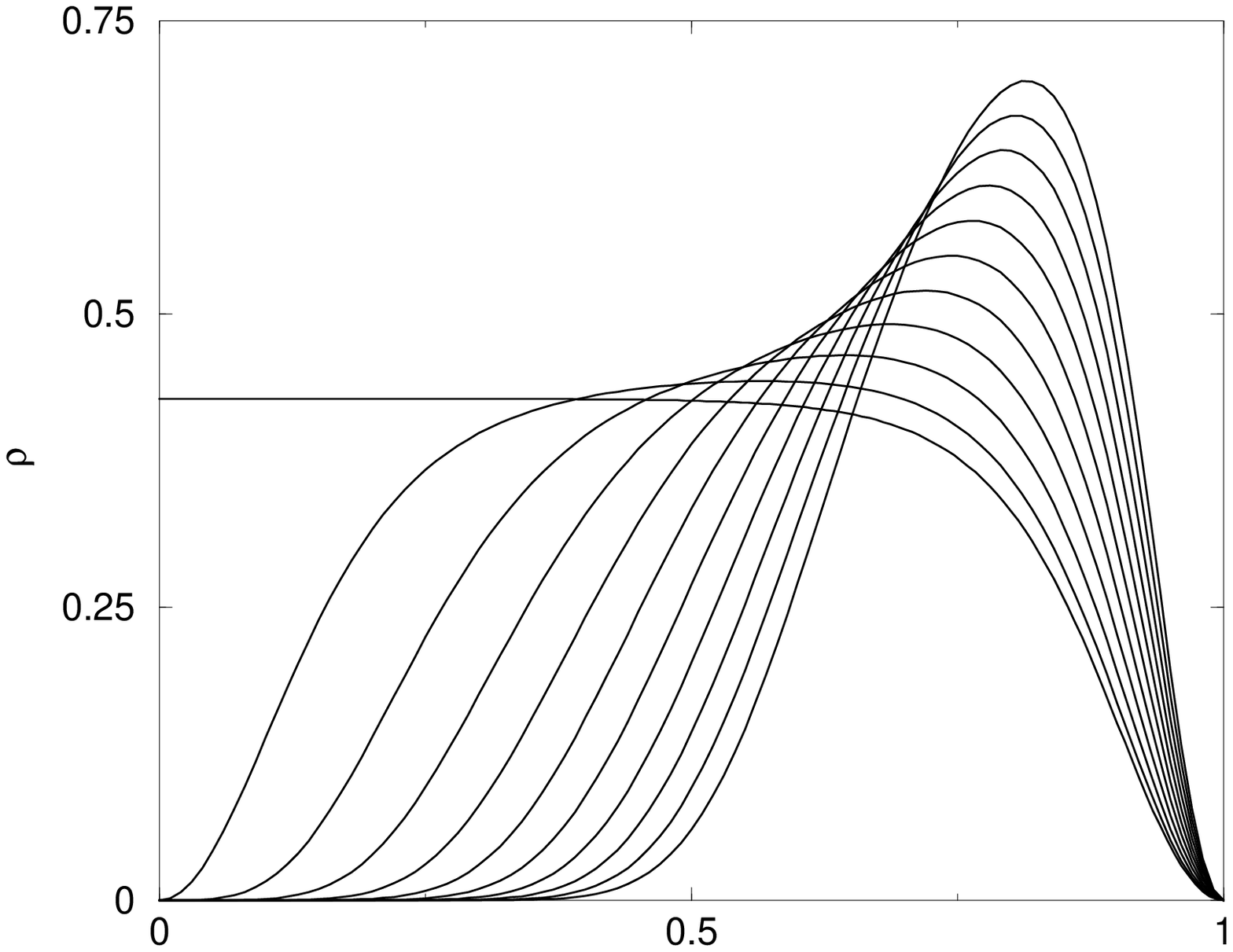} b)\epsfxsize=8truecm
\epsfbox{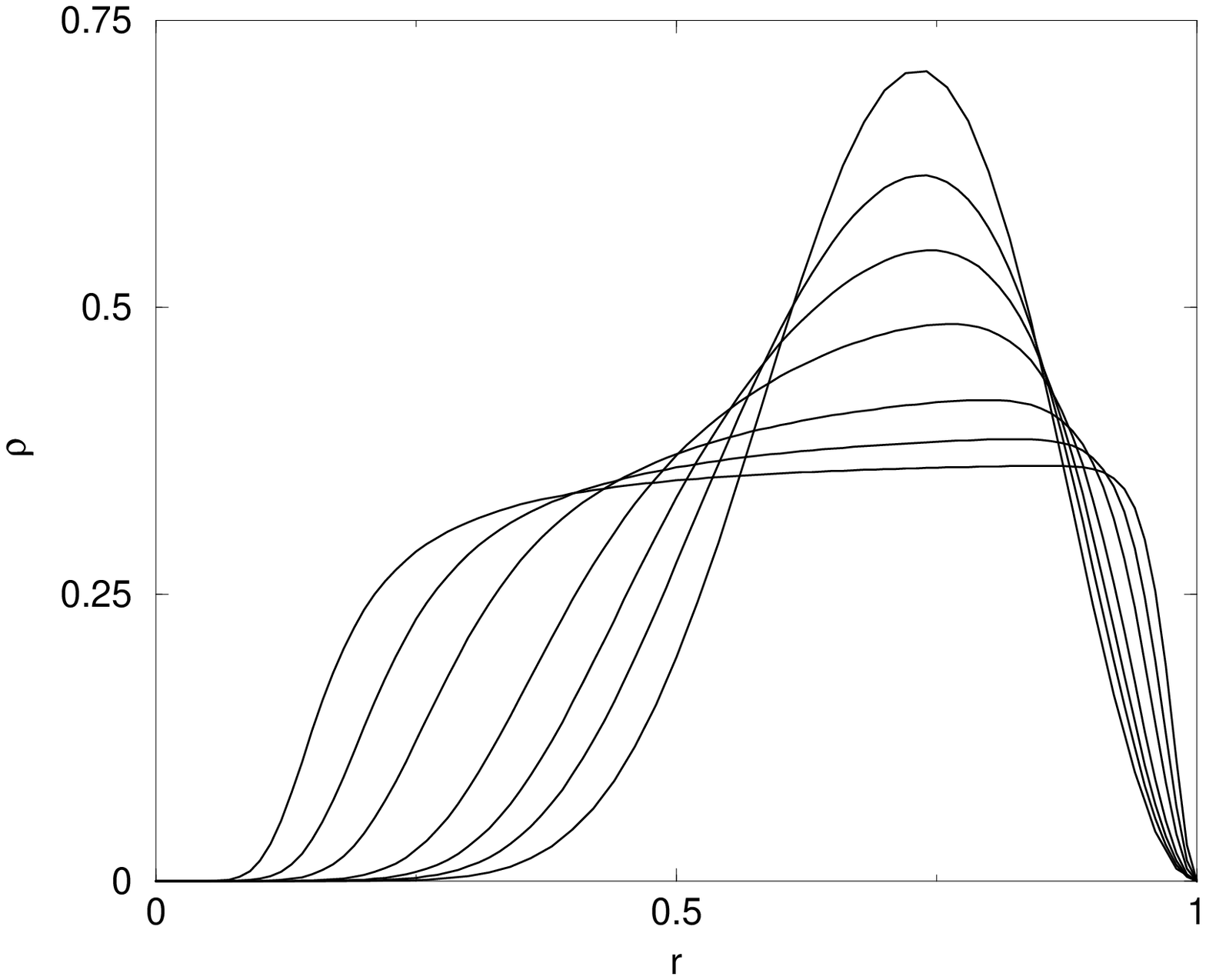}} 
\caption{\protect\small Density of the condensate 
$f_q^2$ for a) $M= 40\pi$ and $q=0,1...10$ b) $q=5$ and $ M/pi= 4,
20, 40, 80, 200$ $400$ and $800$; the higher the maximum of the curve, the smaller 
$ M$ and the bigger $q$. The profiles have been obtained as the stationnary 
solution of equation (\ref{GL1D}) and a regular grid of only $100$ mesh 
points is needed for convergence.
\label{axivort}}
\end{figure}

Where $T_q(M)=\pi \int r \cdot dr((f_q')^2+\frac{q^2}{r^2}f_q^2)$ is the 
kinetic energy and $U_q(M)=M\pi \int r \cdot dr f_q^4$ the nonlinear energy 
term. $E_q(M)=T_q(M)+U_q(M)$ is the total energy of the vortex solution.
The $q$-vortex function $f_q$ is found as before through the dissipative 
dynamics of the axisymetric Ginzburg-Landau equation:

\begin{equation}
\partial_t f_q = \frac{1}{2} \left(f_q''+\frac{f_q'}{r}-\frac{q^2}{r^2}f_q 
\right) - M f_q^3 + \mu_q^0 f_q
\label{GL1D}
\end{equation}
\noindent
so that the equilibrium solutions $f_q$ are local mimina of the free energy 
$F_q$. Figure 
(\ref{axivort} a)) shows the density  $f_q^2$ of the 
solution for $M= 40 \pi$ and $q=0,1,...,10$ as a function of the position 
$r$. As it can be emphasized from 
equation (\ref{eqaxi}), the bigger $q$, the larger is the core of the 
vortex, and the higher is the density level outside the vortex core. 
Indeed, equation 
(\ref{eqaxi}) for $r \rightarrow 0$ implies that the solution $f_q$ behaves 
like $r^q$ near the center. Figure (\ref{axivort} b))
represents also the $q=5$ solution for $M$ varying between $4 \pi$ and 
$800 \pi$. The vortex core is shrinking as $M$ increases in agreement with 
the evolution of the condensate healing length ($\propto 1/\sqrt{M}$),
so does the wall boundary layer.

\begin{figure}
\centerline{\epsfxsize=12truecm \epsfbox{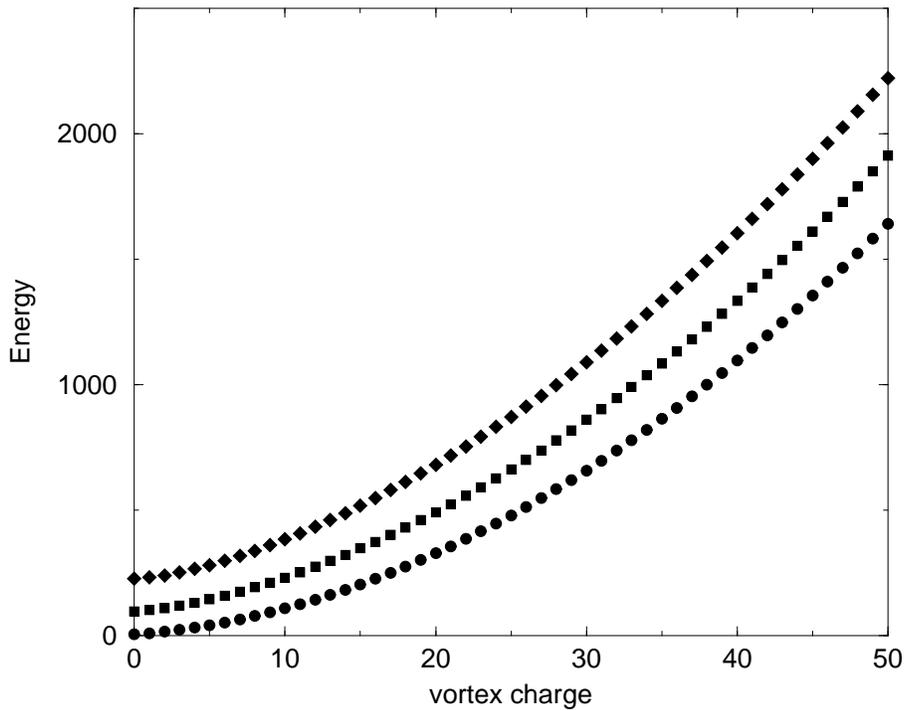}} 
\caption{\protect\small Energy of the vortex solutions as a function of the 
charge $q$ for three different values of $M$. From bottom to top, $M=1$
(circles), $M=40$ (squares) and $M=100$ (diamonds).
\label{energy}}
\end{figure}

Figure (\ref{energy}) shows the free energy of the giant
vortices at $\Omega=0$ as a function of $q$ for different values of 
$M$. The energy is found convex with $q$ for each curve. From this curve
we can determine for each $\Omega$ which $q$-vortex minimizes the free 
energy. It defines an increasing
set of critical frequencies $\Omega_{q}^m(M)$ such that the $q$-vortex is the 
one with the lowest energy for $\Omega_{q-1}^m(M)< \Omega < \Omega_{q}^m(M)$ 
with

$$ \Omega_q^m(M)= E_{q+1}-E_{q} $$

A $q$-vortex is stable when it is a local minimum of the free energy 
(\ref{fener}). We expect from the numerics that such a vortex is unstable
at low enough frequency $\Omega$: it decomposes itself into $q$ single 
vortices (see figures (\ref{quar0}) and (\ref{snap0})). On the other hand,
high frequencies destabilize the $q$-vortex by nucleating new vortices which 
gather into a new giant vortex as observed for $M=100$. Thus, the $q$-vortex
can only be stable in a finite range of frequency and depending on
the vaues of $q$ and $M$, we may observe that the $q$-vortex can never be
stabilized (it is the case for $q=4$, $M=100$).
The stability analysis is performed by an expansion of the
free energy around the $q$-vortex solution.
We write the perturbed wave-function of the giant vortex for $q>1$ in the 
form:
$$ \psi=\frac{\psi_q^0+\delta \psi}{||\psi_q^0+\delta \psi^q||} $$

Following the periodic dependance in $\theta$, we decompose the 
perturbation $\delta \psi^q$ among azimuthal modes as a Fourier set 
$$\delta \psi^q(r,\theta)=\sum_{q'} 
\epsilon_{q'} g_{q'}^q(r)e^{i q' \theta}.$$ 
The complex functions $ g_{q'}^q(r)$
describe the $q'$ pertubation mode of the $q$ vortex and
are normalized to unity. The amplitudes $\epsilon_{q'}$ of the $q'$ 
perturbation modes are complex numbers. We can omitt the $q$ perturbation
mode which will bring nothing to the stability analysis since $\psi_q^0$ is
already the minimizer of the axisymmetrical free energy with charge $q$.
By analysing the perturbed wavefunction
near $r=0$, we observe that for $q'<q$ the perturbation mode describes the
destabilization of the $q$ vortex into a $q'$-vortex in the center surrounded 
by $q-q'$ single vortices. On the other hand, for $q'>q$ the perturbation
describes the nucleation from the wall of $q'-q$ isolated vortices.
We obtain for the perturbed wavefunction:
$$ \psi=\frac{\psi_q^0+\sum_{q'\neq q} \epsilon_{q'} g_{q'}^q(r)e^{i q' \theta}}
{\sqrt{1+\sum_{q' \ne q} |\epsilon_{q'}|^2}} $$

The free energy correction $\delta F$ writes at the first non-zero 
order in the $\epsilon$'s:

$$ \delta F=\sum_{q'<q''} (\epsilon_{q'},\epsilon_{q''}^*) {\cal L}_{q',q''}
 \begin{pmatrix} \epsilon_{q'}^* \\ \epsilon_{q''} \end{pmatrix} $$

omitting the $q$ superscript from now on. The set of operators ${\cal L}_{q',q''}$ follows:

$${\cal L}_{q',q''}= \delta_{q'+q'',2q} \begin{pmatrix} \delta F_{q'} & 
2 \pi M \int rf_q(r)^2g_{q'}(r)g_{q''}(r)dr \\  2 \pi M 
\int rf_q(r)^2g_{q'}^*(r)g_{q''}^*(r)dr & \delta F_{q''} \end{pmatrix}$$

with the diagonal terms:

\begin{equation}
\delta F_{q'}=-T_{q}-2U_{q}+(q-q')\Omega+2\pi \int rdr
\left(\frac{1}{2}((g_{q'}')^2+\frac{(q')^2}{r^2}g_{q'}^2)+2M \cdot g_{q'}^2 f_q^2 \right)
\label{diag}
\end{equation}
 
We notice here that the $\theta$ dependance of the free energy has been 
integrated so that further calculations will only concern the $r$ spatial
variations. This axisymetrical simplification leads eventually to a one 
dimensional stability problem. 
The stability of the $q$-giant follows from the perturbated energy 
structure: if the operators ${\cal L}_{q',q''}$ are all found positive then
the vortex is stable. On the other hand, if at least one of these operators is shown 
negative the giant vortex is unstable towards either decomposition into 
simpler vortices or nucleation/annihilation of vortices. The operators
${\cal L}_{q',q''}$ couple in fact the two modes $g_{q'} $ and $g_{2q-q'} $ 
so that the determination of their full spectrum remains a complicate task. 

However, if we could neglect the non-diagonal terms, the determination of
the giant vortex stability would be straightforward. Indeed, the 
stability would be only determined by the minimization of the effective free 
energies $ \delta F_{q'}$ associated to the functions $g_{q'}$ for each 
$q' \ne q$.  Following this assumption discussed in more details below, we only
have to determine the eigenfunction $\hat{g}_{q'}$ which minimizes the 
corresponding energy:

\begin{equation}
\delta E_{q,q'}=2\pi \int rdr\left(\frac{1}{2}((g_{q'}')^2+\frac{(q')^2}{r^2}g_{q'}^2)
+2M \cdot g_{q'}^2 f_q^2\right)
\label{pener}
\end{equation}

with the following conditions
(omitting the hat of the functions $\hat{g}_{q'}$ later on):

\begin{equation}
<g_{q'}|g_{q'}>=1. \; \; g_0'(0)=0 \;\; g_{q'\ne 0}(0)=0 \;\; g_{q'}(1)=0
\label{restr}
\end{equation}

We observe that energy (\ref{pener}) is unchanged through the symetry 
$q' \rightarrow -q'$ so that we just need to calculate the minimizer for
$q' \ge 0$. 
$g_{q'}$ corresponds in fact to the calculation of a $q'$ giant vortex placed
into the effective potential:

$$ U_{eff}(r)=2 M f_q(r)^2 +V(r) $$

\begin{figure}
\centerline{\epsfxsize=8truecm \epsfbox{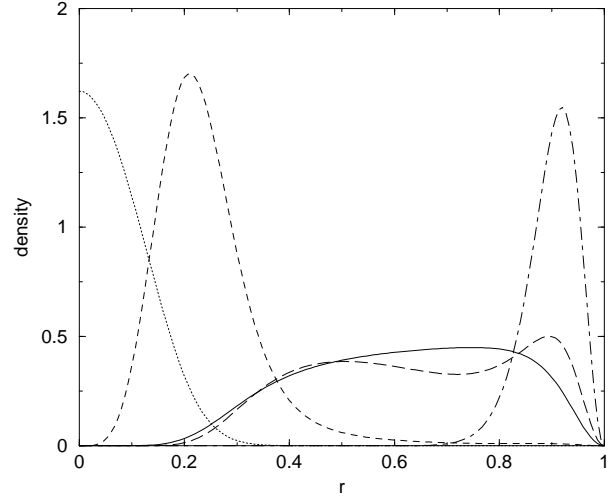}} 
\caption{\protect\small Density profiles of the perturbations modes $g_{q'}$ 
for a $q=4$ giant vortex at $M=300$. Only the profiles for $q'=0$ (dotted
line), $q'=2$ (dashed line), $q'=6$ (long dashed) and $q'=20$ (dot-dashed) are
indicated. The $q=4$ vortex profiles is also shown (solid line). The density
for $q'=0$ and $q'=2$ have been divided by height and two respectively 
for simplicity.
\label{profgq}}
\end{figure}

The solution can be reached again through a dissipative dynamics:

\begin{equation}
\partial_t g_{q'} = \frac{1}{2} \left(g_{q'}''+\frac{g_{q'}'}{r}-\frac{q^2}{r^2}g_{q'}
\right) - 2M f_q^2 g_{q'} + \mu' g_{q'}
\label{GLpert}
\end{equation}
where $\mu'$ is introduced here as the Lagrangian multiplier for the unity
norm constraint.
Figure (\ref{profgq}) shows precisely different $g_{q'}$ for $q=4$ and
$M=300$. The $q$ giant vortex profile is also indicated. As suggested by the 
effective potential, the perturbations modes are concentrated near the 
center of the trap for $q'<q$ where the giant vortex density is very small.
For $q'>q$ the density $g_{q'}$ is concentrated near the wall since the 
centrifugal term of the perturbation dominates the $q$-vortex influence.
Moreover, this figure is in agreement with our assumption of weak off-diagonal
terms made above. Indeed, for $|q'|<q$ the interaction term $ g(q')g(2q-q')$ is
very small since one of the solution is concentrated near $r\sim 0$ while the 
other one is only present near $r=1$. However, for $ q'< -q$ the situation 
cannot be concluded easily since both term is concentrated near $r=1$.

By minimizing the energy $\delta E_{q,q'}$ (\ref{pener}), we can
determine whether the functionnal $\delta F_{q'}$ (\ref{diag}) is positive
definite for all type of perturbations as function of the frequency
$\Omega$. The sign of this free energy for the $q'$-perturbation changes for:

$$ \Omega_c(q')=\frac{T_q+2U_q-\delta E_{q,q'}}{q-q'} $$

\begin{figure}
\centerline{a)\epsfxsize=8truecm \epsfbox{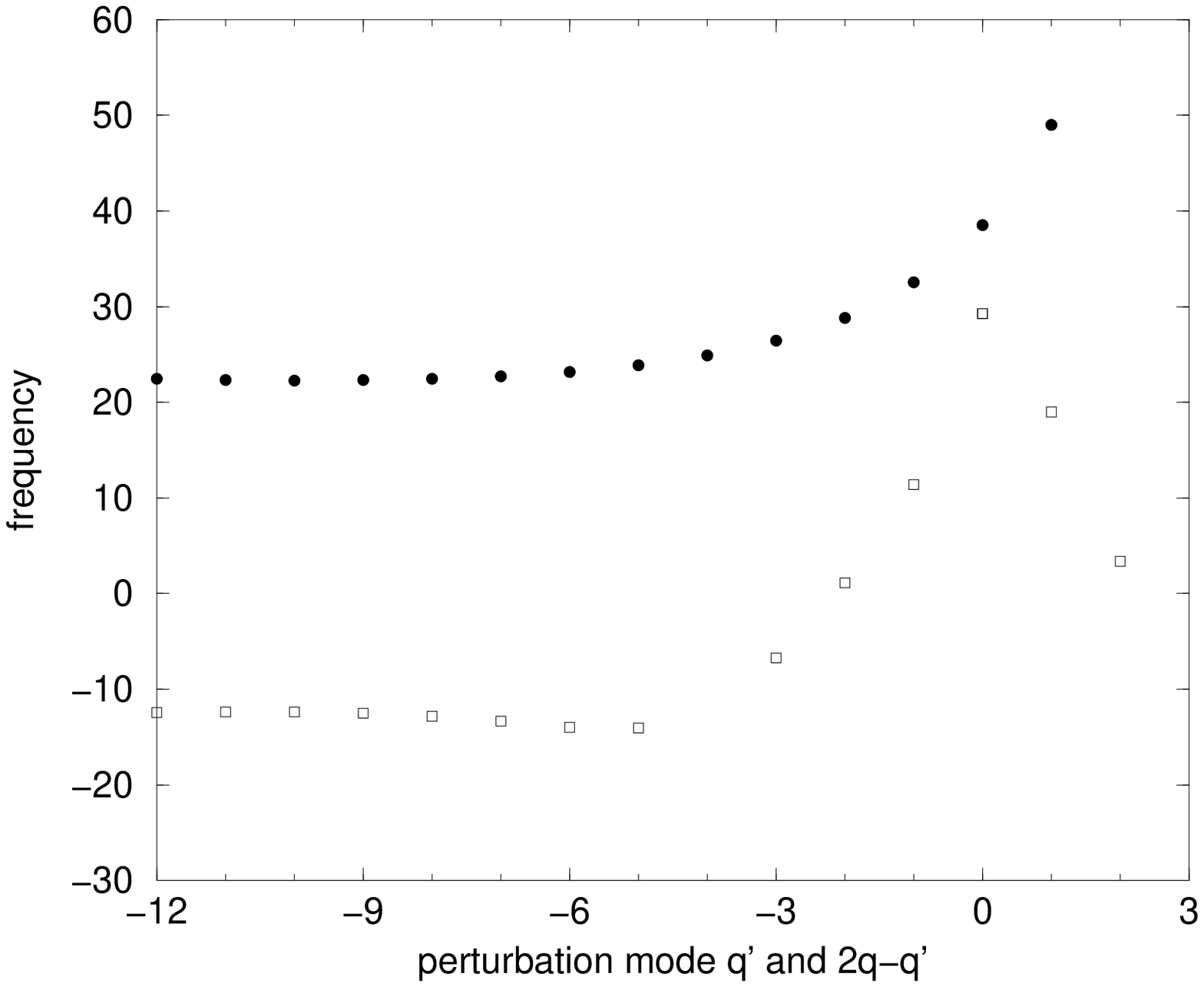} b)\epsfxsize=8truecm \epsfbox{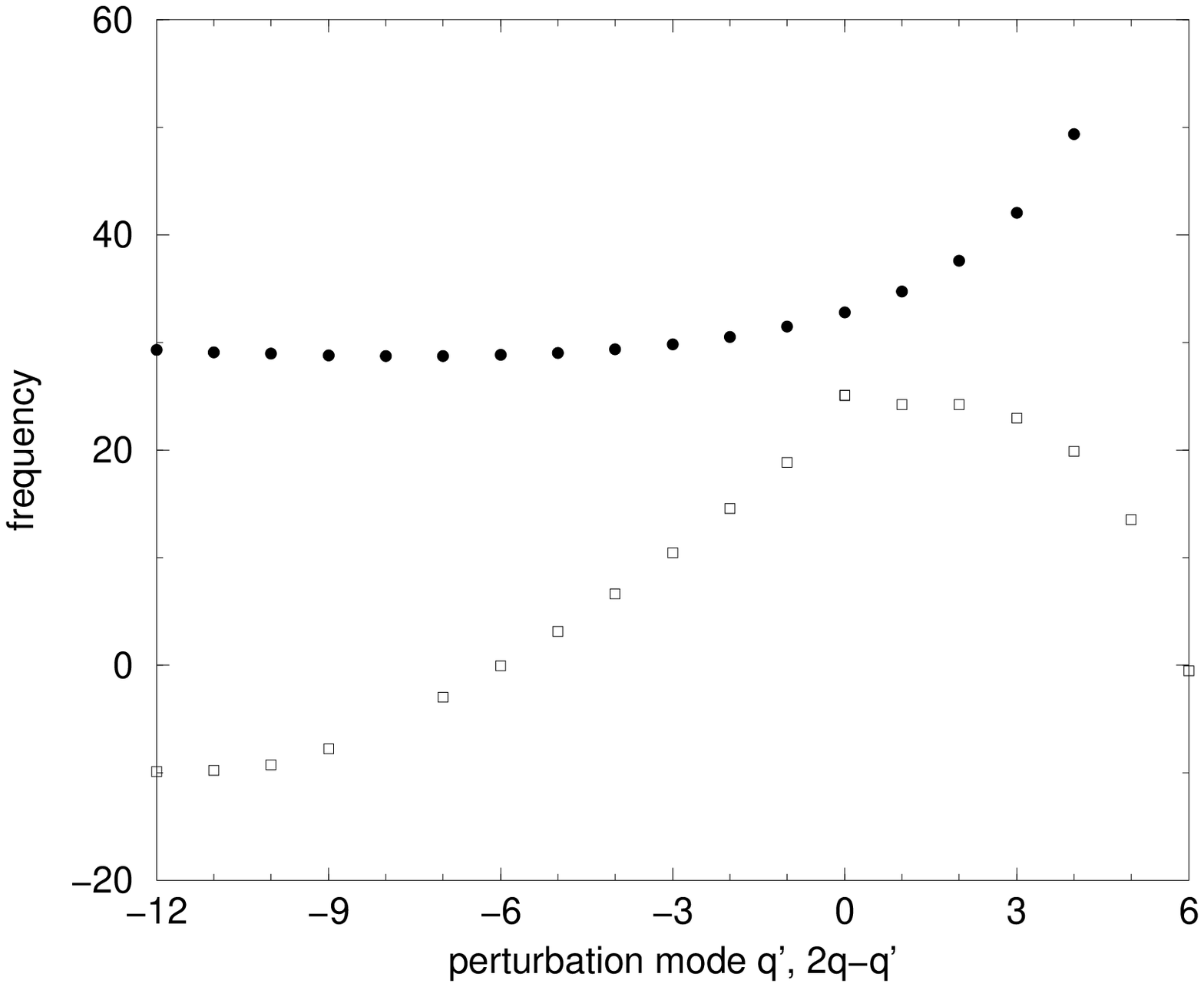}} 
\caption{\protect\small $\Omega_{high}(q')$ (black circles) and $\Omega_{low}$
(squares) as functions of the mode number $q'$ for $M=300$, a) $q=4$, and b) 
$q=8$. It illustrates the two generic situations encountered 
for the stability of a $q$-giant vortex: a) the $4$-giant vortex is never 
stable; b) the $8$-giant vortex is stable for $\Omega$ in the range 
$[25.09,28.76]$.
\label{omeg300}}
\end{figure}

It tells us that for $q'<q$ the $q'$ perturbation mode is stable 
for $\Omega >  \Omega_c(q')$ while for $q'>q$ it is stable only for 
$\Omega < \Omega_c(q')$. The change of inequality is due to the change of
sign of $q-q'$. We define then two critical curves $ \Omega_{low}(q')=
\Omega_c(q')$ and $ \Omega_{high}(q')=\Omega_c(2q-q')$ calculated both for 
$q' <q$. For a given $\Omega$, if one can find a $q'$ such that $ \Omega_{low}
(q')>\Omega$ then the $q$-vortex cannot exist since it decomposes into 
more complex vortices configuration. On the other hand if one can find a 
$q'$ such that $ \Omega_{high}(q')<\Omega$, the $q$ vortex is also unstable 
and new vortices enter in the cloud. Thus the $q$-vortex can only be observed
if ${\rm Max}_{q'<q} \Omega_{low}(q') < {\rm Min}_{q'<q} \Omega_{high}(q')$
and in the range ${\rm Max}_{q'<q} \Omega_{low}(q')< \Omega < {\rm Min}_{q'<q} 
\Omega_{high}(q')$ the $q$-vortex is linearly stable.
Figures (\ref{omeg300}) a) and b) show  $ \Omega_{low}(q')$ and 
$\Omega_{high}(q')$ for $M=300$, $q=4$ and $q=8$ respectively.
This figure is found in relatively good agreement with the full numerical
simulations of the dynamics shown figure (\ref{snap0}).
We remark on figure (\ref{omeg300}) that the $4$-vortex cannot be stabilized 
since the perturbation mode $q'=0$ is still unstable when the high vortices 
perturbation $q'=18$ becomes unstable (at $\Omega=22.29$) as observed
on figure (\ref{snap0}). The solution with four single vortices becomes 
unstable at $\Omega=$ which is close to the value found for the $4$-vortex. 
For $q=8$.....

\begin{figure}
\centerline{\epsfxsize=8truecm \epsfbox{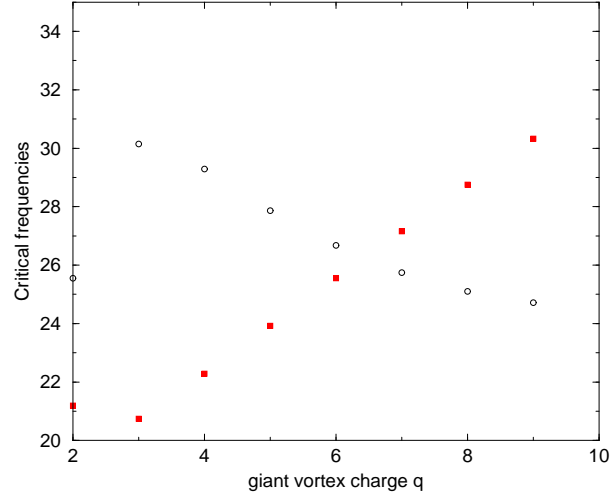}} 
\caption{\protect\small Evolution of ${\rm Max}_{q'<q} \Omega_{low}(q')$ 
(circles) and ${\rm Min}_{q'<q} \Omega_{high}(q')$ (red squares) for the giant
vortex charge $q$ between $2$ and $15$. The two curves cross between $q=6$ and
$q=7$, so that only above $ \Omega = 25.74$ can giant vortices be observed. 
\label{omlh300}}
\end{figure}

Figure (\ref{omlh300}) presents the evolution of ${\rm Max}_{q'<q} 
\Omega_{low}(q')< \Omega < {\rm Min}_{q'<q} \Omega_{high}(q')$ for $q \in
[2,15]$. As observed above, it shows that for $q<7$ no giant vortex can be
observed while for $ \Omega > 25.74$ giant vortices of charge $q>6$ can be
observed in a given range of $\Omega$.

\begin{figure}
\centerline{\epsfxsize=8truecm \epsfbox{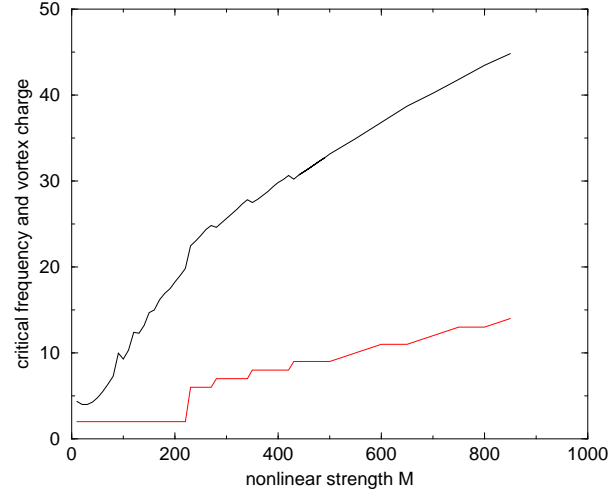}} 
\caption{\protect\small Critical curves $\Omega_g(M)$ (black line) and 
$q_g(M)$ (red line). Giant vortices are linearly stable above these curves.
\label{omgiant}}
\end{figure}

Defining $\Omega_g(M)$ (respectively $q_g(M)$) as the lowest frequency 
(vortex charge) at which a giant vortex can be observed for a strength $M$, we 
determine in the $\Omega$-$M$ plane (or $q$-$M$ plane similarly) the region
where the solution consist in stable giant vortices only. The curves 
$\Omega_g(M)$ and $q_g(M)$ delimiting these regions are shown on figure 
(\ref{omgiant}). 

\section{Discussion}
We have shown numerically that giant vortex solutions can exist in 
axisymmetrical 2-D BEC model for trap potentials steeper than harmonic.
We have investigated in more details the linear stability of these vortices
for the cylindrical wall-like potential. The axisymmetry property of the 
$q$-vortices simplifies the stability procedure into a one dimensional
minimization problem. Under a no off-diagonal terms assumption presenting good 
agreement with the full numerical studies we can determine the regions where 
the giant vortices are stable. Finally a master curve presenting the 
frequency and vortex charge above which giant vortices are stable for varying
nonlinear strength $M$ is presented. These results have been obtained for the
particular case of cylindrical wall potential but can be adapted to
smooth BEC potentials. The stability analysis lead to the same minimization
problem but the numerical
determination of the $q'$ modes needs more careful investigations since they 
are not concentrated on $r \le 1$ disk.

\section{Acknowledgements}
It is my pleasure to thank Vincent Hakim for his encouraging remarks on this
work. I also would like thank warmly the editors for organising this 
special issue. I cannot also finish without express my special thanks to 
Yves Pomeau who made me discover the magic world of nonlinear 
dynamics!

\end{document}